# *Enhancing LTE-Advanced Relay Deployments via Relay Cell Extension*

Ömer Bulakci[1,2], Abdallah Bou Saleh[1,2], Simone Redana[1], Bernhard Raaf[1], Jyri Hämäläinen[2]
1. Nokia Siemens Networks, St.-Martin-Strasse 76, 81541, Munich, Germany
{omer.bulakci, abdallah.bousaleh}ieee.org, {simone.redana, bernhard.raaf}@nsn.com
2. Aalto University School of Electrical Engineering, FI-00076, Espoo, Finland
jyri.hamalainen@aalto.fi

*Abstract*-- **Relaying is a promising enhancement to current radio access networks. Relay enhanced LTE-Advanced networks are expected to fulfill the demanding coverage and capacity requirements in a cost-efficient way. However, due to low transmit power, the coverage areas of the relay nodes will be small. Therefore, the performance of relay deployments may be limited by load imbalances. In this study, we present a practical solution for this problem by introducing a bias to cell selection and handover decisions along with a reduction in eNB transmit power. This method results in an extension of the relay cells and an appropriate load balance can then be achieved. Moreover, it is shown that a proper power control setting is necessary in the uplink and that power control optimization can further enhance the system performance. Comprehensive system level simulations confirm that the proposed solution yields significant user throughput gains both in the uplink and the downlink.**

*Index terms*-- **cell selection, load balancing, LTE-Advanced, power control, relay deployment**

## I. INTRODUCTION

Deploying decode-and-forward relay nodes (RNs) is a promising solution for Long Term Evolution (LTE)-Advanced networks to meet the growing demand and challenging requirements for coverage extension and capacity enhancement [1][2]. RNs have been recently agreed to be standardized during the 3GPP LTE-Advanced work item phase for the coverage-improvement scenario [3]. RNs are relatively small nodes with low power consumption and connect to the core network with a wireless backhaul through an evolved Node B (eNB). This feature enables deployment flexibility and eliminates the high costs of a fixed backhaul. Besides, RNs do not have strict installation guidelines with respect to radiation and planning regulation. Hence, installing RNs involves lower operational expenditure (OPEX) [4] and faster network upgrade when operators aim to improve quality of service (QoS) [5].

An RN can be considered a wireless eNB, which controls its own cell (Type 1 RN as specified in 3GPP). That is, the RN has its own physical cell ID and includes functionalities such as radio resource management, scheduling and hybrid-ARQ (HARQ) retransmissions [6]. In line with the LTE context, cell selection can be done conventionally according to the received signal strength in the downlink (DL). The lower transmit power of an RN will then render a smaller coverage area relative to that of an eNB and hence, a lower cell load. Thus, the available resources in the RN cell might not be fully exploited, whereas in the eNB cell, the competition for the resources is still high.

In this study, a combination of reduction in eNB transmit power and biasing in cell selection and handover decisions is proposed to balance the loads between eNB and RN cells. Moreover, an LTE-compliant power control is optimized in the uplink (UL). Note that biasing is backward compatible with LTE Rel. 8. Thus, the legacy LTE Rel. 8 terminals can also support this load balancing approach.

The rest of the paper is organized as follows. The concept of RN cell extension is introduced in Section II. Section III presents the system model and simulation assumptions. In Section IV thorough performance evaluation and analysis are provided. The paper is concluded in Section V.

## II. RN CELL EXTENSION

Fig. 1 shows the DL received power levels from the eNB and RN given at different distances from the eNB. The RN coverage area is defined by the point of intersection of the received power levels from the RN and the eNB, i.e. when the received powers from RN and eNB are equal at the user equipment (UE). The imbalance in coverage areas between RN and macro cells is attributed to the RN low transmit power, low antenna gains and high path-loss exponent, since RN antennas are typically placed under the rooftop. Such a coverage imbalance might lead to load imbalance as well.

To tackle this problem, a combination of eNB transmit power reduction and biasing in favor of RNs in cell selection and handover decisions is proposed. As illustrated in Fig. 1, an X dB reduction in eNB transmit power will reflect in X dB reduction in the UE received signal power (from the eNB), therefore, increasing the RN range. Added to that, a Y dB bias in thresholds for cell selection and handover decisions will further extend RN range.

The concept of biasing in the DL leads to a trade-off: The signal-to-interference-plus-noise-ratio (SINR) of the cell edge UEs decreases but there will be more time-frequency resources available for newly adopted UEs, since the number of users in the RN cell is lower than that in the eNB cell. Note that the allowable SINR degradation is limited by a pre-defined lower bound to ensure a reliable transmission. Commonly this bound is -7 dB defining the constraint on biasing.

Besides, the eNB power reduction will increase the SINR of UEs newly adopted by the RNs. Yet, UEs connected to the eNB will experience a lower SINR. This may increase the outage which sets a practical limit on the eNB power reduction. In this study, the upper limit on power reduction is 16 dB.

It is crucial to notice that even a larger imbalance will occur on the UL side. As depicted in Fig. 2, due to the higher transmit power of the eNB, UE2 will connect to the eNB, although it has a higher path-loss towards the eNB than towards the RN. Because of the UL power control, UE2 will then transmit at a higher power level and increased interference levels will be experienced in the UL. Thus, RN cell extension is expected to considerably improve the UL performance. Note that X dB reduction in eNB transmit power along with Y dB biasing corresponds to X+Y dB *effective biasing* in the UL.

### III. SYSTEM MODEL

The simulated network has a regular hexagonal cellular layout with 19 tri-sectored sites. The RNs admit regular outdoor deployment at the sector border. Fig. 3 presents example deployments of 4 RNs and 10 RNs per sector. Simulation parameters follow the parameter settings agreed in 3GPP [6] and are summarized in Table I.

Ten uniformly distributed indoor UEs are dropped per sector and full buffer traffic model is applied. Full reuse scheme and a round robin (RR) scheduler are considered. All available resources in a cell are assumed to be used by its UE(s).

SINR is mapped to link throughput using Shannon approximation as explained in [7]. In the considered two-hop case, UEs are either served directly by an eNB or indirectly via an RN. As the relay link (eNB-RN link) is modeled as an ideal link, the end-to-end throughput is equal to that on the RN-UE link. Shadowing is considered only for non-line-of-sight (NLOS) connections, while fast fading is not simulated.

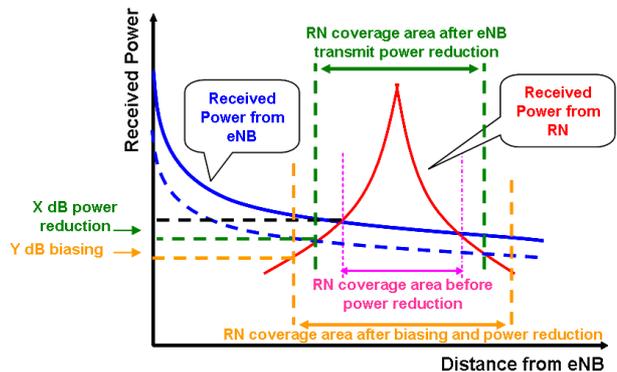

Figure 1. Illustration of RN cell extension concept.

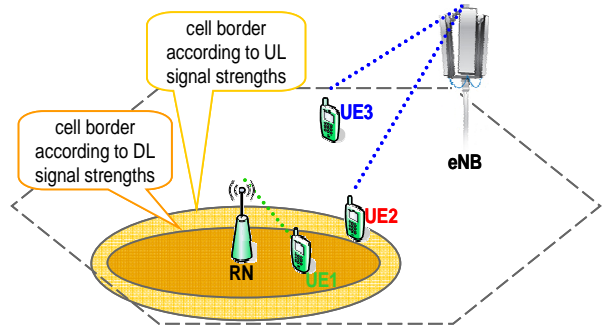

Figure 2. The coverage imbalance between DL and UL.

Power control is applied only in the UL. The LTE Rel. 8 compliant open-loop power control scheme, as investigated in [8], is applied. Note that power control is an important means to adjust the receiver dynamic range [8], where a high receiver dynamic range increases the susceptibility of Single Carrier- Frequency Division Multiple Access (SC-FDMA) to the loss of orthogonality and thus, can cause intra-cell interference [9]. The transmit power of a UE is given in (1).

$$P = \min\{P_{\max}, P_0 + 10 \cdot \log_{10} M + \alpha \cdot L\} \text{ [dBm]} \quad (1)$$

Here $P_{\max}$ is the maximum allowed transmit power of the UE, $P_0$ is a cell-specific parameter used for controlling the SNR target and it ranges from -126 dBm to 23 dBm ($P_{\max}$) with a step size of 1 dB and $M$ is the number of physical resource blocks (PRBs) allocated to one UE, where the PRB defines the resource allocation granularity. Furthermore, $\alpha$ is the cell-specific path loss compensation factor that can be set to 0 or range from 0.4 to 1.0 with a step size of 0.1, and $L$ is the DL path loss estimate. The parameters $P_0$ and $\alpha$ can be optimized according to the desired performance optimization strategy [8].

## IV. PERFORMANCE EVALUATION

Urban and suburban scenarios with inter-site-distances (ISDs) of 500 m and 1732 m are considered [6]. Results are provided for both the UL and DL. The eNB-only deployment is considered as a reference to determine the performance gains.

Simulation results are obtained assuming an ideal relay link. We note that the throughput cumulative distribution functions (CDFs) for relay and pico eNB deployments (with independent backhaul) are almost the same for low CDF percentiles. At higher percentiles, due to the limited throughput on the relay link, the gain for relay deployments is reduced compared to pico eNB deployments. In our study, we focus on the coverage-improvement capabilities of RNs and consider the 5%-ile throughput CDF level as a decisive criterion. As the relay link is not expected to be the bottleneck for such levels, results are assumed to reflect also the performance of a system with non-ideal relay link.

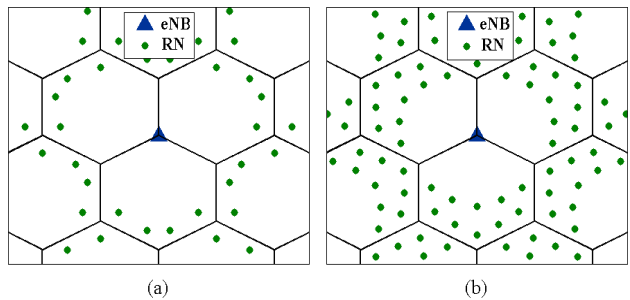

Figure 3. RN deployments; (a) 4 RNs and (b) 10 RNs.

TABLE I. SIMULATION PARAMETERS

| Parameter | Value |
|---|---|
| **System Parameters** | |
| Carrier Frequency | 2 GHz |
| Bandwidth | 10 MHz |
| Highest MCS | 64-QAM, R = 9/10 |
| Penetration Loss | 20 dB for UEs only |
| Thermal Noise PSD | -174 dBm/Hz |
| SINR lower bound | -7 dB |
| **eNB Parameters** | |
| eNB Transmit Power | 46 dBm |
| eNB Elevation Gain | 14 dBi |
| eNB Antenna Configuration | Tx-2, Rx-2 |
| **UE Parameters** | |
| Maximum Transmit Power | 23 dBm |
| UE Antenna Configuration | Tx-1, Rx-2 |
| **Relay Node Parameters** | |
| RN Transmit Power | 30 dBm |
| RN Antenna Configuration | Tx-2, Rx-2 |
| RN-UE Elevation Gain | 5 dBi |
| **Channel Models** | |
| Based on | TR 36.814 v9.0.0 [6] |
| **Shadowing** | |
| Shadow Fading | Log-normal |
| Standard Deviation | 8 dB (eNB-UE), 10 dB (RN-UE) |

### A. Urban Scenario (ISD 500 m)

RN cell extension is first analyzed in the DL. Fig. 4 presents the relative gain in UE throughput at the 5%-ile CDF level for different combinations of RN biasing and eNB transmit power reductions. The figure illustrates results in a 4-RN deployment only; simulations show similar behavior in a 10-RN deployment. Macro eNB-only deployment is considered as a reference. Fig. 4 shows an increase in the gain for higher power reduction and a fluctuation along the different biasing values. We note that gains are significantly high. Since more UEs are added to the RN cells, macro cell users experience less competition for resources due to off-loading the eNB. As well, UEs newly adopted by the RNs will benefit from the availability of more resources. Hence, such users are able to significantly improve their throughput levels.

The RN cell extensions realized by biasing and power reduction are given in Table II for 4-RN and 10-RN deployments. Results show that the RN area is almost doubled in the former case whereas an increase of 75% is achieved in the latter deployment.

Along the gain from RN cell extension, higher eNB power reduction (within a limit) will decrease the interference on RN cells, thus, improving the performance. It is worth noting that the urban scenario is interference-limited.

The optimized resultant gains over the eNB-only deployment at the 5%-ile throughput CDF levels are respectively 264% and 520% compared to 71% and 148% that are achieved by non-optimized 4-RN and 10-RN deployments (see Table II). These gains are achieved with 2 dB biasing and 16 dB power reduction in 4-RN deployments and 1 dB biasing and 16 dB power reduction in 10-RN deployments.

Fig. 5 presents the UE throughput CDF plots for 4-RN and 10-RN deployments with and without RN cell extension optimizations. The eNB-only scheme is shown for reference. The throughput CDF plots again reflect the significant gain from balancing loads among RN and macro cells. It is worth pointing out that load balanced 4-RN deployments may outperform default 10-RN deployments.

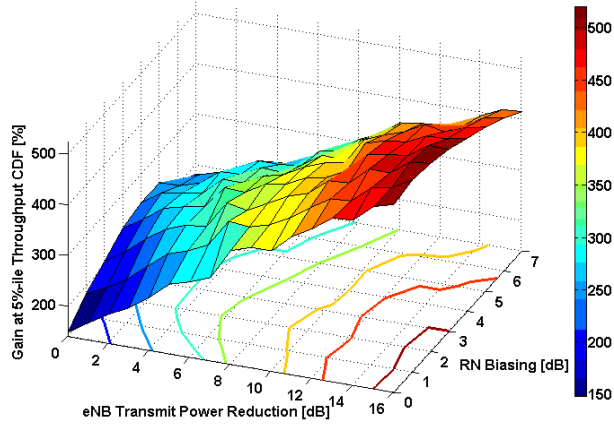

Figure 4. Relative UE throughput gain due to biasing and eNB power reduction. 4-RN deployment, DL urban scenario.

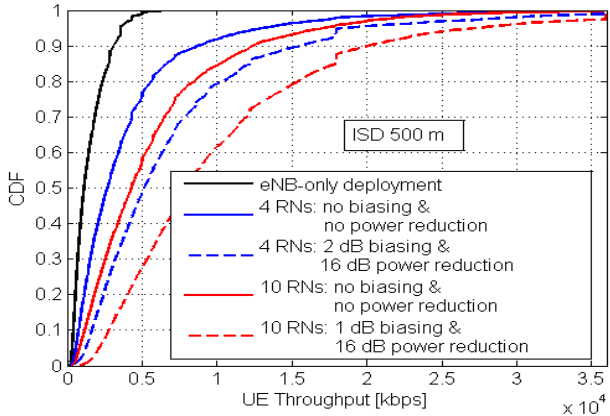

Figure 5. DL throughput CDFs. 4-RN deployment, urban scenario.

TABLE II. DL GAINS IN URBAN SCENARIO, ISD 500 M

| RN Bias [dB] Reference: eNB-only deployment | 5%-ile throughput gain [%] | | RNs coverage area [%] | |
|---|---|---|---|---|
| | 4 RNs | 10 RNs | 4 RNs | 10 RNs |
| No bias | 71 | 148 | 29 | 44.8 |
| Optimum bias and power reduction | 264 | 520 | 61 | 77 |

The UL performance is analyzed next. In [8], the optimum parameter settings are investigated both for eNB-only and RN deployments. Taking this study as a basis, three different power control optimization strategies can be applied:
  i. The eNB-only setting at all nodes [8]
  ii. Optimized setting in RN deployment [8]
  iii. Optimized settings in RN deployment for each biasing value.

These optimization strategies are performed such that, the 5%-ile UE throughput is maximized without degrading the 50%-ile UE throughput.

Fig. 6 presents the 5%-ile UE throughput gains vs. different biasing values, where the eNB-only deployment is taken as the reference. It can be seen that, via a proper power control optimization, biasing can significantly improve the performance of RN deployments. As well, optimum performance is obtained over the entire range of the investigated biasing values when power control optimization is done for each biasing value (solid curves in Fig. 6). On the other hand, if eNB-only setting is applied, the performance degrades rapidly at high biases, since this strategy does not adapt to conditions imposed by biasing. The maximum gains of 373% and 708% are observed when 21 dB and 18 dB biasing values are applied for the 4-RN and 10-RN deployments, respectively. The corresponding power control settings are given in Table III.

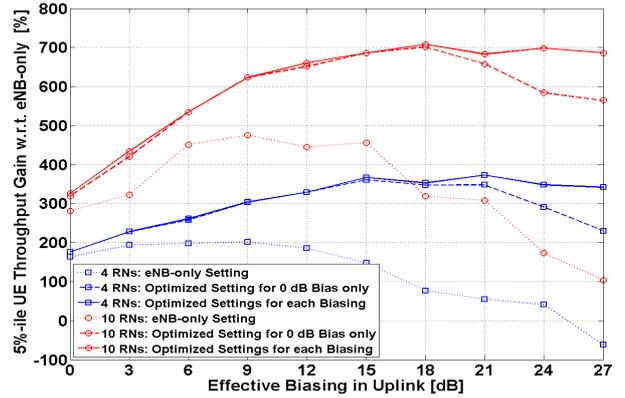

Figure 6. 5%-ile UE throughput gains for 4-RN and 10-RN deployments, UL urban scenario.

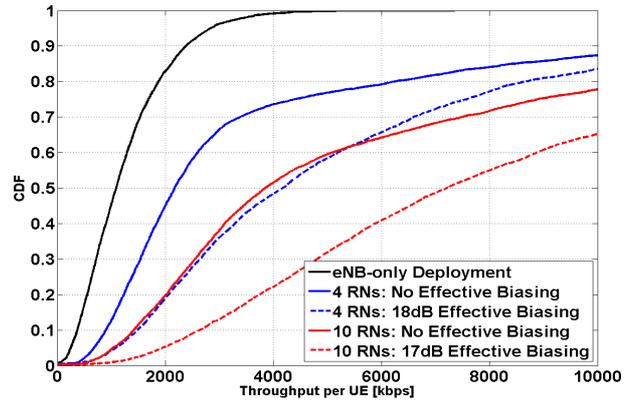

Figure 7. UE throughput CDFs for 4-RN and 10-RN deployments with DL-limited effective biasing values, UL urban scenario.

TABLE III. UL POWER CONTROL PARAMETER CONFIGURATIONS IN URBAN SCENARIO, ISD 500 M

| Parameters | | Strategy I | | Strategy II | | Strategy III | |
|---|---|---|---|---|---|---|---|
| | | eNB | RNs | eNBs | RNs | eNBs | RNs |
| $P_0$ | 4 RNs | -101 | -101 | -95 | -101 | -91 | -101 |
| | 10 RNs | -101 | -101 | -95 | -101 | -93 | -101 |
| $\alpha$ | | 1.0 | 1.0 | 1.0 | 1.0 | 1.0 | 1.0 |
| $P_{max}$ | | 23 | 23 | 23 | 15 | 23 | 15 |

Different biasing values have been derived in the DL and the UL. However, since a UE is connected to the same eNB or RN in both, DL limits the valid biasing in UL to 18 dB and 17 dB for 4-RN and 10-RN deployments, respectively. The resultant UE throughput CDFs are presented in Fig. 7.

## B. Suburban Scenario (ISD 1732 m)

The suburban scenario is coverage-limited, i.e. system performance degrades in case of any transmission power reduction. While in ISD 500 m scenario, we have seen an increase in system performance when decreasing eNB transmission power, Fig. 8 shows a different behavior. A decrease in eNB power causes a significant decrease in SINRs of coverage-limited UEs and deteriorates the performance. Hence, it is seen that RN cell extension is best achieved with biasing the cell selection and handover thresholds.

The optimum 5%-ile throughput CDF levels are achieved for 4 dB biasing and no power reduction in 4-RN deployments. This yields 221% gain over eNB-only deployments, while 4-RN deployments without any RN cell extension techniques achieve a 171% gain (see Table IV). In 10-RN deployments, the highest gain is 639% and is achieved with 2 dB RN biasing and 4 dB power reduction. 10-RN deployments with default cell selection achieve a significantly high gain of 578%. It is worth noting that when deploying 10 RNs per sector, the system becomes slightly interference-limited and hence, moderate power reduction becomes viable again.

As concluded from Table IV, the relative gains due to balancing cell loads via RN cell extension in ISD 1732 m scenario are moderate. This is as well reflected in the low RN cell extensions which are around 8% and 5% in 4-RN and 10-RN deployments, respectively. That is, a proper load balancing is not attained. Such a behavior can be attributed to the already large RN cells and hence, the relatively high resource competition within. Another reason is the suburban path loss model having very good propagation conditions to the eNB up to some distance after which they start deteriorating fast. When the RNs are first deployed, they already cover a large part of the sector and it is hard to further expand into the area with good propagation conditions to the eNB without suffering from high eNB interference or applying large power reductions which will deteriorate the macro UEs.

Fig. 9 presents the UE throughput CDFs. The modest gains from RN cell extension are again seen in the throughput distributions.

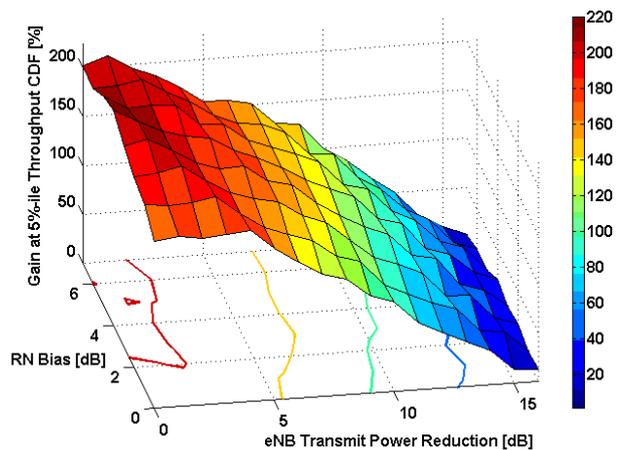

Figure 8. Relative throughput gain due to biasing and eNB power reduction. 4-RN deployment, DL suburban scenario.

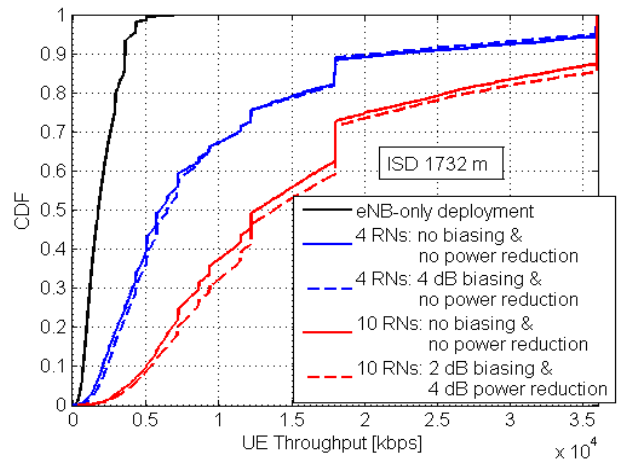

Figure 9. DL throughput CDFs. 4-RN deployment, suburban scenario.

TABLE IV. DL GAINS IN SUBURBAN SCENARIO, ISD 1732 M

| RN Bias [dB] Reference: eNB-only deployment | 5%-ile throughput gain [%] | | RNs coverage area [%] | |
|---|---|---|---|---|
| | 4 RNs | 10 RNs | 4 RNs | 10 RNs |
| No bias | 172 | 578 | 43 | 67.3 |
| Optimum bias and power reduction | 221 | 639 | 46.5 | 71 |

For the UL performance analysis, power control optimization strategies as presented for urban scenario are also applied for the suburban scenario where the study in [10] is taken as a basis. It is seen in Fig. 10 that the different strategies yield similar performances. Moreover, the gains due the biasing are smaller compared to the urban scenario. The maximum gains of 321% and 1947% can be achieved for 4-RN and 10-RN deployments when 6 dB and 12 dB biasing values are applied, respectively. The corresponding power control parameter settings are given in Table V. Moreover, too high biasing yields less gain due to the decrease in SINR values of macro UEs and the increased

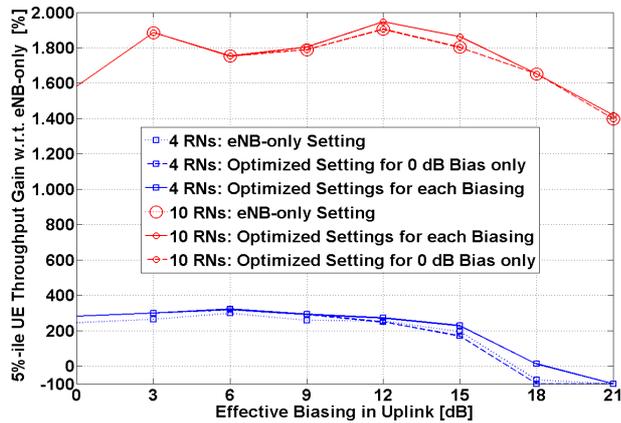

Figure 10. 5%-ile UE throughput gains for 4-RN and 10-RN deployments, UL suburban scenario.

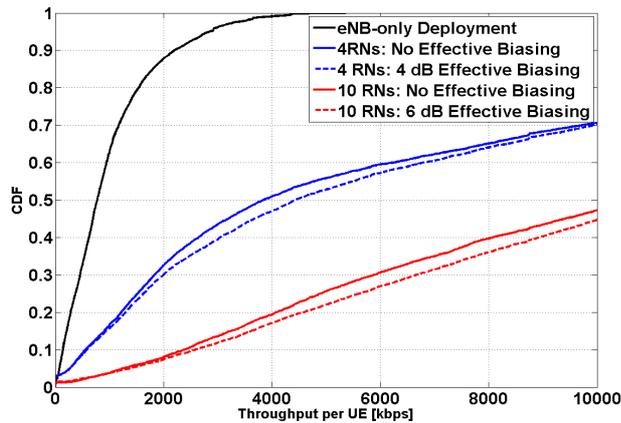

Figure 11. UE throughput CDFs for 4-RN and 10-RN deployments with DL-limited effective biasing values, UL suburban scenario.

TABLE V. UL POWER CONTROL PARAMETER CONFIGURATIONS IN SUBURBAN SCENARIO, ISD 1732 M

| Parameters | | Strategy I | | Strategy II | | Strategy III | |
|---|---|---|---|---|---|---|---|
| | | eNB | RNs | eNBs | RNs | eNBs | RNs |
| $P_0$ | 4 RNs | -63 | -63 | -67 | -63 | -65 | -63 |
| | 10 RNs | -63 | -63 | -63 | -63 | -61 | -63 |
| $\alpha$ | | 0.6 | 0.6 | 0.6 | 0.6 | 0.6 | 0.6 |
| $P_{max}$ | | 23 | 23 | 23 | 23 | 23 | 23 |

resource competition among relay UEs, which cannot be mitigated by power control optimization.

Similar to the urban scenario, the DL limits effective biasing in UL to 4 dB and 6 dB for 4-RN and 10-RN deployments, respectively. The resultant UE throughput CDFs are presented in Fig. 11, showing that biasing mostly improves higher percentiles.

## V. CONCLUSIONS

Relay node deployments offer a viable solution to achieve considerable coverage improvements in the uplink and downlink, both in 3GPP urban and suburban scenarios. The performance of the relay node deployment can be further enhanced by load balancing between macro cells and relay cells. In this paper, we have proposed to use a combination of eNB transmission power reduction and biasing cell selection and handover thresholds as a practical solution, which is compatible with existing LTE Rel. 8 specifications.

Following the 3GPP guidelines, we have used the 5%-ile throughput CDF level as a performance measure in simulations. Results show that a high eNB transmission power reduction along with moderate biasing of cell selection and handover thresholds can achieve significantly higher gains on the downlink compared to non-optimized relay node deployments in urban scenario. However, optimization gains in the suburban scenario are modest relative to the huge gain already achieved with non-optimized relay node deployments.

Significant improvements, similar to those on the downlink, are also experienced on the uplink in the urban scenario, given that a proper power control setting is utilized. The gains are again modest for the suburban scenario.